# Ultra-clean interface between high *k* dielectric and 2D MoS$_2$

Han Yan[1+], Yan Wang[1+*], Yang Li[1], Dibya Phuyal[2], Lixin Liu[1], Hailing Guo[3], Yuzheng Guo[3], Tien-Lin Lee[4], Min Hyuk Kim[5], Hu Young Jeong[5] & Manish Chhowalla[1*]

**Affiliations:**

[1]*Department of Materials Science & Metallurgy, University of Cambridge, 27 Charles Babbage Road, Cambridge CB3 0FS, UK*

[2]*Division of Material and Nano Physics, Department of Applied Physics, KTH Royal Institute of Technology, Stockholm, SE-106 91, Sweden*

[3]*School of Electrical Engineering and Automation, Wuhan University, Wuhan, China*

[4]*Diamond Light Source, Harwell Science and Innovation Campus, Didcot OX11 0DE, United Kingdom*

[5]*UNIST Central Research Facilities (UCRF) and School of Materials Science and Engineering, UNIST, Ulsan, South Korea.*

[+]*These authors contributed equally to this work.*

*Correspondence to: yw472@cam.ac.uk, mc209@cam.ac.uk

**Abstract:** Atomically thin (or two-dimensional, 2D) transition metal dichalcogenide (TMD) semiconductors possess ideal attributes for meeting industry scaling targets for transistor channel technology. The realization of ultra-scaled field effect transistors (FETs) will require industry compatible gate dielectrics with very low equivalent oxide thickness (EOT) values. Dielectric substrates such as SiO$_2$, Al$_2$O$_3$ and HfO$_2$ unintentionally dope 2D TMDs and create interfacial defect states that lead to non-ideal FET characteristics – such as variable threshold voltage. Despite their importance, the electronic properties of 2D TMD/dielectric interface have not been extensively studied. Here we show that zirconium oxide (ZrO$_2$) – a well-known industry compatible high dielectric constant (*k*) oxide – forms an ultra-clean interface with MoS$_2$. Our detailed soft and hard X-ray photoelectron spectroscopy (XPS) analysis reveals that SiO$_2$ and HfO$_2$ substrates introduce significant doping of MoS$_2$ while ZrO$_2$ exhibits no measurable interactions with $MoS_2$ – consistent with density functional theory calculations. Because of the ultra-clean $MoS_2$/$ZrO_2$ dielectric interface, back gated monolayer $MoS_2$ FETs using $ZrO_2$ as dielectric show extremely stable and positive threshold voltage of 0.36 ± 0.3 V, low subthreshold swing of 75 mV/dec, and high ON currents of > 400 µA. We also demonstrate P-type 2D $WSe_2$ FETs with high ON state currents of > 200 µA/µm via suppression of electron doping using $ZrO_2$ dielectrics. Atomic resolution imaging and of $ZrO_2$ deposited on top of $MoS_2$ also reveals a defect free interface, which enables top gate FETs with equivalent oxide thickness of 0.86 nm and SS values of 80 mV/dec. Furthermore, we demonstrate that the ultraclean interface between $ZrO_2$ and monolayer $MoS_2$ enables effective modulation of the threshold voltage in top gate FETs by gate metal work function engineering. Our results show that $ZrO_2$ holds tremendous promise as an industry compatible high k dielectric for electronics based 2D TMD semiconductors.

**Introduction:**

2D TMD semiconductors hold tremendous promise for next generation of sub-10 nm channel FETs [1–3]. The source-drain current and switching speed of FETs depends on the gate capacitance. Therefore, gate dielectrics with high dielectric constants ($k$) that can be scaled to very low equivalent oxide thicknesses (EOTs) are desirable. In bulk semiconductors, covalent bonding between the semiconductor and dielectric leads to low defect concentration at the interface, which minimizes threshold voltage instabilities [4]. 2D TMDs form weak van der Waals bonds with dielectrics. Therefore, surface defect states on dielectrics remain unpassivated and can lead to charge transfer, interface states, and electrostatic doping from dipoles. This results in highly variable 2D TMD FETs with substantial swings in threshold voltage, low ON state currents, and low subthreshold slopes. Therefore, pristine interfaces between 2D TMDs and dielectrics are essential for high performance FETs.

Key requirements for dielectrics are that they must be high $k$ to enable continued scaling, have large band offsets (> 1 eV) to prevent injection into its bands, have minimal electrically active defects, and form a clean interface with 2D TMDs [4]. Despite their importance, suitable dielectrics that are complementary metal oxide semiconductor (CMOS) compatible for 2D TMD FETs have yet to be identified. Studies show that commonly used oxide dielectrics such as $SiO_2$, $Al_2O_3$, and $HfO_2$ heavily dope $MoS_2$ [5–7], leading to N-type FETs operating in depletion mode. P-type FETs suffer from low ON state hole currents because of electron doping from the dielectrics [8,9]. These effects are due to interface hybridization between sulfur-oxygen at the $MoS_2/SiO_2$ interface and the formation of new chemical bonds like Hf-S when $MoS_2$ is in contact with oxygen deficient $HfO_x$ surface [10,11]. These defects at the $MoS_2$/dielectric interface leads to enlarged hysteresis and increased subthreshold swing along with large threshold voltage variation in FET transfer characteristics.

The most common inert dielectric substrate used by academic researchers for 2D TMD devices is hexagonal boron nitride, but it has a low dielectric constant [12]. Recent work has shown that it is possible to minimize the interactions between the dielectric and 2D TMDs by making free standing $SrTiO_3$ membranes on which $MoS_2$ is placed [13]. $Y_2O_3$ thin film transfer has also been reported [14]. A single-crystalline $Al_2O_3$ dielectric, formed via oxidation, has been reported for top-gate $MoS_2$ FETs [15]. In these cases, the dry transfer of the dielectrics leads to van der Waals gap between the insulator and semiconductor to minimize interactions and form clean contacts [13–15]. Similarly, it has been shown that increasing the van der Waals gap between $HfO_2$ and $MoS_2$ can decouple interactions at the interface [11].

In this work, we have examined the interface interactions between 2D MoS$_2$ and three most common industry compatible dielectrics: SiO$_2$, HfO$_2$, and ZrO$_2$ using synchrotron photoelectron spectroscopy. To ensure that interfaces are not influenced by extrinsic artefacts, we employed UV ozone-cleaned PDMS as the medium for transfer of CVD-grown MoS$_2$, which results in clean samples that are free of residue and well adhered to the substrate (**See Methods**). A typical AFM image of the samples used in this study and the corresponding height profile are shown in **Fig. 1a,** with larger scale of the AFM image in **Extended Data Figure 1.** The measured thickness of 0.7 nm matches almost exactly to that of single layer MoS$_2$, suggesting that the samples are free of residue at the interface between the dielectric and MoS$_2$.

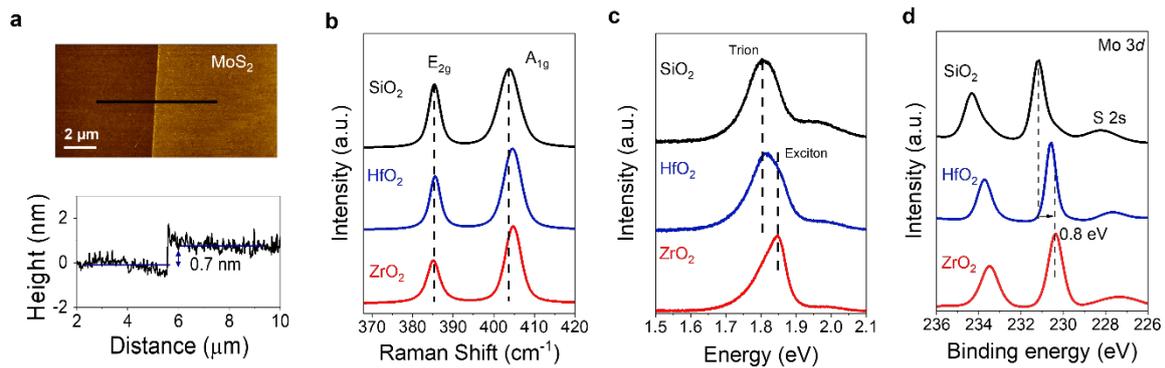

**Fig. 1: Monolayer MoS$_2$ on SiO$_2$, HfO$_2$ and ZrO$_2$ dielectric substrates. a**, AFM image (top) and height profile (bottom) of transferred MoS$_2$. **b**, Raman spectra of MoS$_2$ on SiO$_2$, HfO$_2$ and ZrO$_2$. The redshift of the A$_{1g}$ mode of MoS$_2$ on HfO$_2$ and SiO$_2$ indicates electron doping from substrate. **c**, PL spectra of MoS$_2$ transferred on SiO$_2$, HfO$_2$ and ZrO$_2$ showing exciton dominant peak on ZrO$_2$ and trion dominant peak on SiO$_2$. HfO$_2$ shows both trion and exciton peaks. **d**, Mo 3$d$ peaks of MoS$_2$ measured by XPS on different dielectrics showing Fermi level shift of 0.8 eV on SiO$_2$ and 0.3 eV on HfO$_2$ with respect to ZrO$_2$. MoS$_2$ on SiO$_2$ has higher binding energy due to more electron doping from SiO$_2$.

The influence of dielectric substrates on 2D MoS$_2$ was probed by Raman, photoluminescence (PL), and X-ray photoelectron spectroscopy (XPS), the results of which are shown in **Fig. 1bd**, respectively. Raman spectra show that the E$_{2g}$ peak of MoS$_2$ remains relatively constant on different dielectrics while the A$_{1g}$ peak of MoS$_2$ on SiO$_2$ and HfO$_2$ exhibits broadening and redshift compared to ZrO$_2$, suggesting electron doping of MoS$_2$ from SiO$_2$ and HfO$_2$ [16]. This is consistent with PL spectra of monolayer MoS$_2$ on SiO$_2$ that show trion dominant emission due to doping from the substrate and exciton dominant emission on ZrO$_2$ (**Fig. 1c**). Mo 3$d$ spectra of 2D MoS$_2$ on different dielectrics were collected using soft X-rays (1 keV). The binding energy of Mo 3$d$ from MoS$_2$ on SiO$_2$ is 0.5 eV higher than MoS$_2$ on HfO$_2$ and 0.8 eV higher than MoS$_2$ on ZrO$_2$, indicating Fermi level of MoS$_2$ on SiO$_2$ is 0.8 eV closer to the conduction band compared to MoS$_2$ on ZrO$_2$. Additionally, the Mo 3$d$ and S 1$s$ (intensity is higher than S 2$p$) spectra collected using 3 keV X-rays exhibit the same trend, as shown in **Extended Data Fig. 2**.

To investigate the interface interactions between MoS$_2$ and dielectrics, we collected XPS spectra as a function of depth using a combination of soft and hard X-rays with energies of 1 keV, 3 keV, and 5.9 keV (**Fig. 2a**). We extracted the valence band edge profile from the surface of the dielectric to a depth of ~ 17 nm for SiO$_2$, 10 nm for HfO$_2$ and 11.5 nm for ZrO$_2$ (see **Methods**). The binding energy shifts in XPS are caused by depth-dependent changes to the Fermi level position within the band gap, which can be determined to reconstruct band edge energy profile at the MoS$_2$/dielectric interface. By comparing the MoS$_2$/dielectric interface with pristine dielectric surface, the interactions between monolayer MoS$_2$ and different dielectrics can be probed.

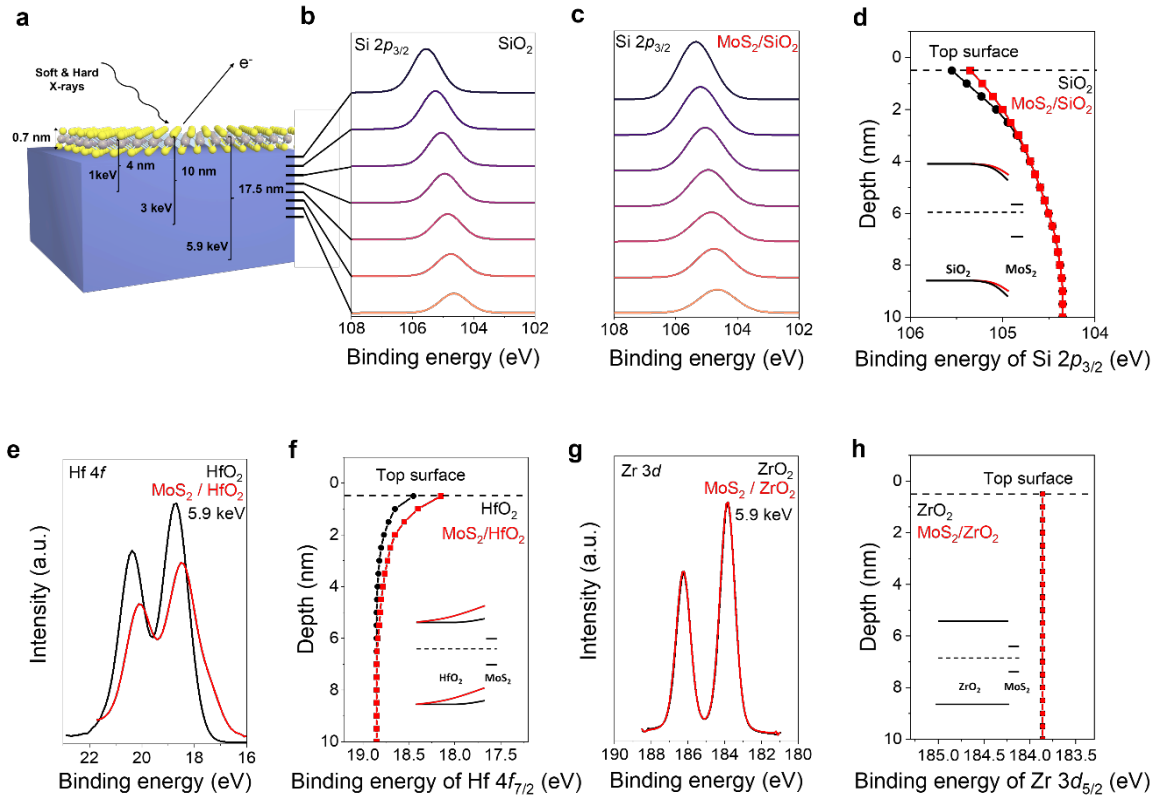

**Fig. 2: MoS$_2$/dielectric interfaces measured by soft and hard X-ray photoelectron spectroscopies. a**, Schematic of the photoelectron setup with probing depths labeled for SiO$_2$. The probing depths for HfO$_2$ are: 2.5 nm at 1 keV, 5.5 nm at 3 keV, and 10 nm at 5.9 keV. The probing depths for ZrO$_2$ are: 3.0 nm at 1 keV, 6.5nm at 3 keV, and 11.5 nm at 5.9 keV. **b**, Depthresolved Si 2$p_{3/2}$ spectra from the SiO$_2$ substrate extracted from the XPS modelling. **c**, Depthresolved Si 2$p_{3/2}$ spectra from the MoS$_2$ on SiO$_2$ sample extracted from the XPS modelling. **d**, Summary of the binding energies of Si 2$p_{3/2}$ at different depths for SiO$_2$ and MoS$_2$ on SiO$_2$. Inset is a schematic of the corresponding SiO$_2$ surface band bending with and without MoS$_2$ on top. **e**, Hf 4$f$ spectra with and without MoS$_2$ measured with 5.9 keV X-rays. **f**, Summary of the binding energies of Hf 4$f_{7/2}$ at different depths derived from the XPS modeling for HfO$_2$ and MoS$_2$ on HfO$_2$. Inset: schematic of the corresponding HfO$_2$ surface band bending before and after contacting MoS$_2$. **g**, Zr 3$d$ spectra with and without MoS$_2$ showing no measurable peak shift or broadening. **h**, Summary of binding energy positions of the Mo 3d$_{5/2}$ at different depths for ZrO$_2$ and MoS$_2$ on ZrO$_2$. Inset shows no band bending at MoS$_2$/ZrO$_2$ interface.

The depth-resolved Si 2$p_{3/2}$ peaks from SiO$_2$ with and without MoS$_2$ on top are shown in **Fig. 2b** and **2c**. The Si 2$p_{3/2}$ peak of pristine SiO$_2$ shifts to higher binding energies as the depth decreases. The Si 2$p_{3/2}$ binding energies with MoS$_2$ on top of SiO$_2$ follow a similar trend. The binding energies of the Si 2$p_{3/2}$ peak as a function of the depth are summarized in **Fig. 2d** for the two cases. It can be seen that the band edge remains constant at depths ≥ 6.5 nm. Whereas significant shifts in the Fermi level (binding energy) are observed at depths < 6.5 nm – causing the band to bend downwards. MoS$_2$ on SiO$_2$ does not influence the shifts in Si 2$p_{3/2}$ binding energy for depths beyond the top ~ 2 nm, where the binding energy shifts are lowered by the presence of MoS$_2$, due to charge transfer at the interface. The inset of **Fig. 2d** shows the energy band diagrams extracted from XPS analysis with depth. It shows that a significant downward band bending of 1.2 eV ± 0.1 eV within the top ~ 6 nm in SiO$_2$, which reduced to 1.0 eV ± 0.1 eV upon the deposition of monolayer MoS$_2$. Hf 4$f$ spectra from HfO$_2$ and MoS$_2$/HfO$_2$ are summarized in **Fig. 2e**. The Hf 4$f$ peaks recorded with 5.9 keV X-rays show significant differences, indicating that the influence of MoS$_2$ extends well into HfO$_2$. It can be seen from **Fig. 2f** (and inset) that the band bending for pristine HfO$_2$ is upwards and it increases to ~ 0.4 eV ± 0.1 eV when MoS$_2$ is deposited on top. The strong interface interactions are indicated by the substantial band bending – extending ~ 6 nm into HfO$_2$. A direct comparison of Si 2$p_{3/2}$ and Hf 4$f$ binding is presented in **Extended Data Fig. 3a-f**. These results clearly show that SiO$_2$ and HfO$_2$ strongly perturb the MoS$_2$/dielectric interface.

In contrast, XPS analysis of ZrO$_2$ shows no measurable surface states and deposition of MoS$_2$ on top does not perturb the surface. The XPS spectra of Zr 3$d$ peak measured at 5.9 keV from pristine ZrO$_2$ and with MoS$_2$ on top are shown **Fig. 2g**. The Zr 3$d$ spectra with other probing energies are summarized in **Extended Data Fig. 3g-i.** No measurable band bending is observed with or without $MoS_2$ across all probing energies. The Zr $3d$ binding energy as a function of the depth is shown in **Fig. 2h**. These results indicate that the interface between $MoS_2$ and $ZrO_2$ is pristine without interactions. The local density of states (LDOS) of the $MoS_2/SiO_2$, $MoS_2/HfO_2$ and $MoS_2/ZrO_2$ interfaces were calculated using density functional theory (see **Methods** and **Extended Data Fig. 4**). The results show that both $SiO_2$ and $HfO_2$ form dipoles at interface with $MoS_2$ due to unpassivated oxygen bonds on the $SiO_2$ surface and incomplete reconstruction of $HfO_2$ surface. In contrast, the surface of $ZrO_2$ does not show any states that can cause band bending – consistent with our synchrotron photoelectron spectroscopy results.

To investigate whether the pristine dielectric interface between $ZrO_2$ and $MoS_2$ can be translated into better and reproducible device performance, we performed electrical measurements on FETs fabricated using chemical vapor deposited (CVD) monolayer $MoS_2$ as the channel with 90 nm $SiO_2$, 40 nm $HfO_2$ and 40 nm $ZrO_2$ dielectrics as back gates (**Fig. 3**). The linear and logarithmic transfer characteristics of FETs using different dielectrics are shown in **Fig. 3a** and **3b**, respectively. It can be seen from **Fig. 3a** that FETs with $SiO_2$ and $HfO_2$ dielectrics operate in depletion mode (channel is ON at zero gate voltage) while FETs with $ZrO_2$ work in enhancement mode (channel is OFF at zero gate voltage). The vast majority of monolayer $MoS_2$ FETs reported in literature work in depletion mode in which a conductive channel is already formed at zero gate bias and require relatively high negative gate voltages to turn the devices off. The transfer characteristics of one typical batch of over 50 monolayer $MoS_2$ FETs fabricated with $ZrO_2$ dielectric (**Fig. 3b**) show that these devices completely turn off at zero gate voltage. In contrast, devices with $HfO_2$ and $SiO_2$ dielectric exhibit high current at zero gate voltage, consistent with electron doping observed in the Raman, PL and XPS analysis. Our standard FET testing protocol involves taking forward and backward scans

(**Extended Data Fig. 5**) and ensuring that the gate leakage current is $\ll 10^{-9}$ A. The output curves of monolayer MoS$_2$ FETs with ZrO$_2$ dielectrics are shown in **Fig. 3c**, showing ON currents of 423 µA with channel length of 1 µm and width of 2.5 µm.

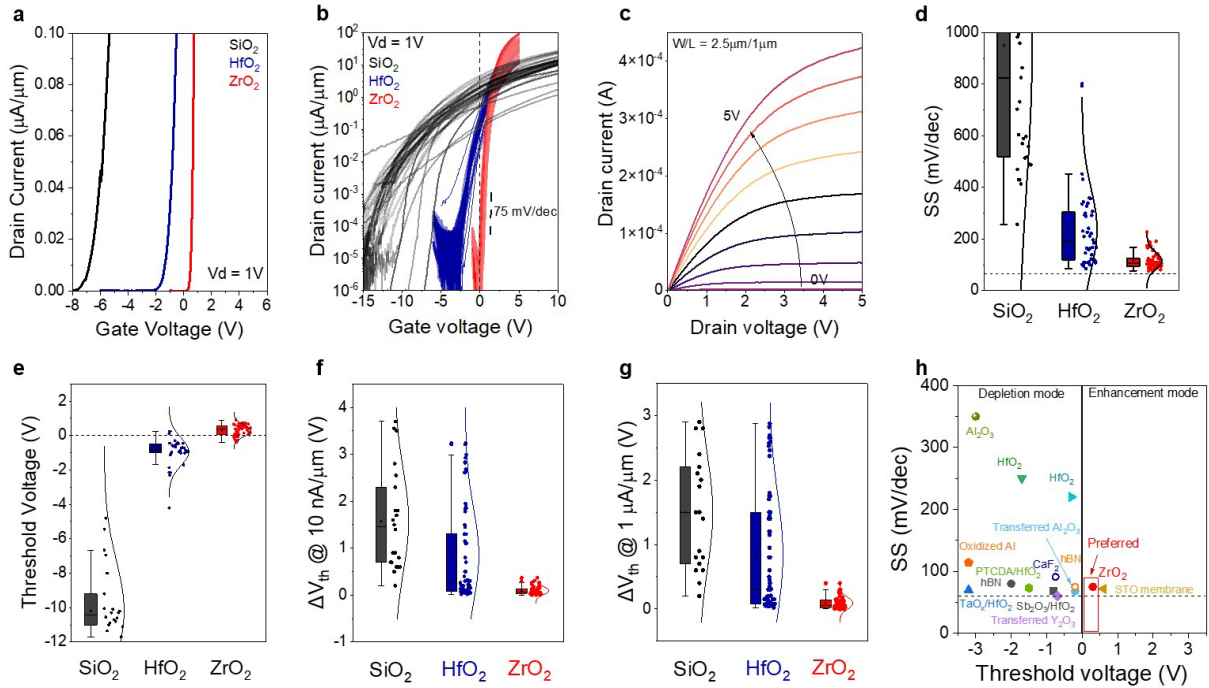

**Fig. 3. MoS$_2$ FET device characteristics with SiO$_2$, HfO$_2$ and ZrO$_2$ dielectrics. a,** Transfer curves of MoS$_2$ FETs plotted linearly to show the negative threshold voltages for MoS$_2$ on SiO$_2$ and HfO$_2$ and positive threshold voltage for MoS$_2$ on ZrO$_2$. **b,** Logarithmic transfer characteristics of MoS$_2$ FETs showing stable and low SS in devices with ZrO$_2$ dielectrics compared to those with SiO$_2$ and HfO$_2$ dielectrics. **c,** Output curves of MoS$_2$ FETs with ZrO$_2$ showing absence of barrier at the contacts and good saturation behaviour. **d-e,** Statistical distribution of SS and V$_{th}$ of the MoS$_2$ FETs. ZrO$_2$ dielectric shows low SS and small positive threshold voltage with a narrow distribution that indicated reproducibility and consistency. **f-g,** Statistical distribution of threshold voltage variations (ΔV$_{th}$) extracted at (f) 10 nA/µm and (g) 1µA/µm for monolayer MoS$_2$ FETs with SiO$_2$, HfO$_2$ and ZrO$_2$ dielectrics. **h,** Summary of

SS and $V_{th}$ from monolayer MoS$_2$ FETs, which shows that most MoS$_2$ FETs devices reported in literature work in depletion mode [13,14,17–27].

The average subthreshold swing (SS) for bottom gated MoS$_2$ FETs with ZrO$_2$ dielectric was 115 ± 32 mV/dec, considerably lower than that of HfO$_2$ (237 ± 157 mV/dec) and SiO$_2$ (950 ± 563 mV/dec), as summarized in **Fig. 3d**. ZrO$_2$-based FETs demonstrated high reproducibility and consistent performance, achieving a minimum SS of ~75 mV/dec. MoS$_2$ FETs incorporating 12 nm HfO$_2$ as bottom dielectrics (**Extended Data Fig. 6 a-b**) demonstrated SS values of 151 ± 32 mV/dec, which still fall short of the performance achieved by devices utilizing 40 nm ZrO$_2$ bottom dielectrics. Moreover, bottom-gated MoS$_2$ FETs with ZrO$_2$ dielectrics scaled down to an EOT of ~ 1 nm demonstrated further improved SS values to 84 ± 14 mV/dec (**Extended Data Fig. 6 c-e**). The highest mobility value of MoS$_2$ FETs with ZrO$_2$ is 114 cm$^2$V$^{-1}$s$^{-1}$, higher than MoS$_2$ FETs with HfO$_2$ (42 cm$^2$V$^{-1}$s$^{-1}$) and SiO$_2$ (26 cm$^2$V$^{-1}$s$^{-1}$). Additionally, multilayer MoS$_2$ FETs were fabricated with SiO$_2$, HfO$_2$, and ZrO$_2$ as dielectrics. The influence of dielectric doping from SiO$_2$ and HfO$_2$ on multilayer MoS$_2$ is also prevalent, as indicated by the transfer curves shown in **Extended Data Fig. 7.**

We extracted the threshold voltage ($V_{th}$) using the constant current method ($I_d$ = 10 nA/μm) during gate voltage sweeps from negative to positive [28]. The $V_{th}$ distribution of MoS$_2$ FETs using SiO$_2$, HfO$_2$ and ZrO$_2$ dielectrics is summarized in **Fig. 3e**. $V_{th}$ for monolayer MoS$_2$ FETs with ZrO$_2$ was found to be 0.36 ± 0.3 V, while with HfO$_2$, it was – 0.93 ± 0.8 V. The $V_{th}$ for MoS$_2$ FETs with SiO$_2$ is even more negative (– 10.2 ± 3.9 V) due to electron doping. The distribution of $V_{th}$ in **Fig. 3e** shows that all of our devices with ZrO$_2$ show a small and positive voltage – indicating that MoS$_2$ FETs with ZrO$_2$ work in enhancement mode as is desirable for digital circuits [29,30]. In contrast, the $V_{th}$ for FETs with SiO$_2$ and HfO$_2$ is variable and negative. The influence of back gate work function on the $V_{th}$ is shown in **Extended Data Fig. 8**.

Hysteresis in FETs was quantified by extracting threshold voltage variations ($\Delta V_{th}$) at 10 nA/μm and 1 μA/μm (**Fig. 3f** and **Fig. 3g**). The results show that FETs with $ZrO_2$ achieve lower hysteresis and improved stability. While progress has been made on realization of low SS FETs (see summary in **Fig. 3h**), here we show for the first time that it is possible to make enhancement mode FETs using industry compatible high $k$ dielectrics.

To demonstrate the general applicability of $ZrO_2$ dielectrics, we have also tested P-type FETs using multilayer $WSe_2$ as the semiconductor channel on different dielectrics. It is well known that electron doping from $SiO_2$ significantly reduces the hole current in $WSe_2$ P-type FETs [8,9] as observed in **Extended Data Fig. 9**. The transfer characteristics of $WSe_2$ FETs with van der Waals Pt contacts using different dielectrics show that the hole current in FETs with $WSe_2$ on $ZrO_2$ (~ 200 μA/μm) is one order of magnitude higher than $WSe_2$ FETs with $HfO_2$ (~ 30 μA/μm) and two orders of magnitude higher than $WSe_2$ FETs with $SiO_2$ (~1.5 μA/μm) – due to suppressed electron doping from $ZrO_2$. The hole current realized on $WSe_2$ FET with $ZrO_2$ dielectric is among the highest reported in the literature, as shown in **Extended Data Fig. 9**.

To demonstrate practical feasibility, we fabricated top gate FETs with $ZrO_2$ dielectric (EOT of 0.86 nm) on multi- and mono-layered $MoS_2$. The top dielectric/semiconductor interface was characterised by cross-sectional scanning transmission electron microscopy (STEM) and Xray photoemission spectroscopy. **Fig. 4a** shows that ~4.0 nm of $ZrO_2$ can be grown on $MoS_2$ without causing damage to the $MoS_2$. Soft and hard X-ray photoemission spectroscopy analyses of 4.0 nm $ZrO_2$ grown on CVD monolayer $MoS_2$ are presented in **Fig. 4b**. The binding energy of Mo $3d_{5/2}$ in the $ZrO_2/MoS_2/ZrO_2$ probed by hard X-rays is around 230.0 ± 0.1 eV, similar to that of as-transferred $MoS_2$ on $ZrO_2$ substrate, indicating the top dielectric deposition does not induce chemical interactions at the interface.

**Fig. 4c** shows typical transfer characteristics of top gate FETs based on multilayer $MoS_2$ with $ZrO_2$ dielectric and Pd/Au gate metal. The results show negligible hysteresis, low subthreshold voltage, and SS of ~ 80 mV/dec. Transfer characteristics top gate FETs on CVD monolayer $MoS_2$ with different gate metals are shown in **Fig. 4d** (and **Extended Date Fig. 10**). The most important results we obtain with top gate devices is that we demonstrate that the threshold voltages with $ZrO_2$ top gate dielectric can be tuned from negative to positive. In contrast, FETs based on monolayer $MoS_2$ with $HfO_2$ as the top dielectric suffer from significantly higher SS values (292 ± 106 mV/dec), large hysteresis, and negative threshold voltages due to the defective dielectric/$MoS_2$ interface (**Fig. 4e** and **Extended Data Fig. 10**). The clean $ZrO_2$ on $MoS_2$ interface also enables effective $V_{th}$ modulation using gate metals with different work functions, as summarised in **Fig. 4f**. These results highlight that the semiconductor/dielectric interface plays a vital role in enabling reliable threshold voltage modulation, minimising hysteresis, and achieving low subthreshold swing values.

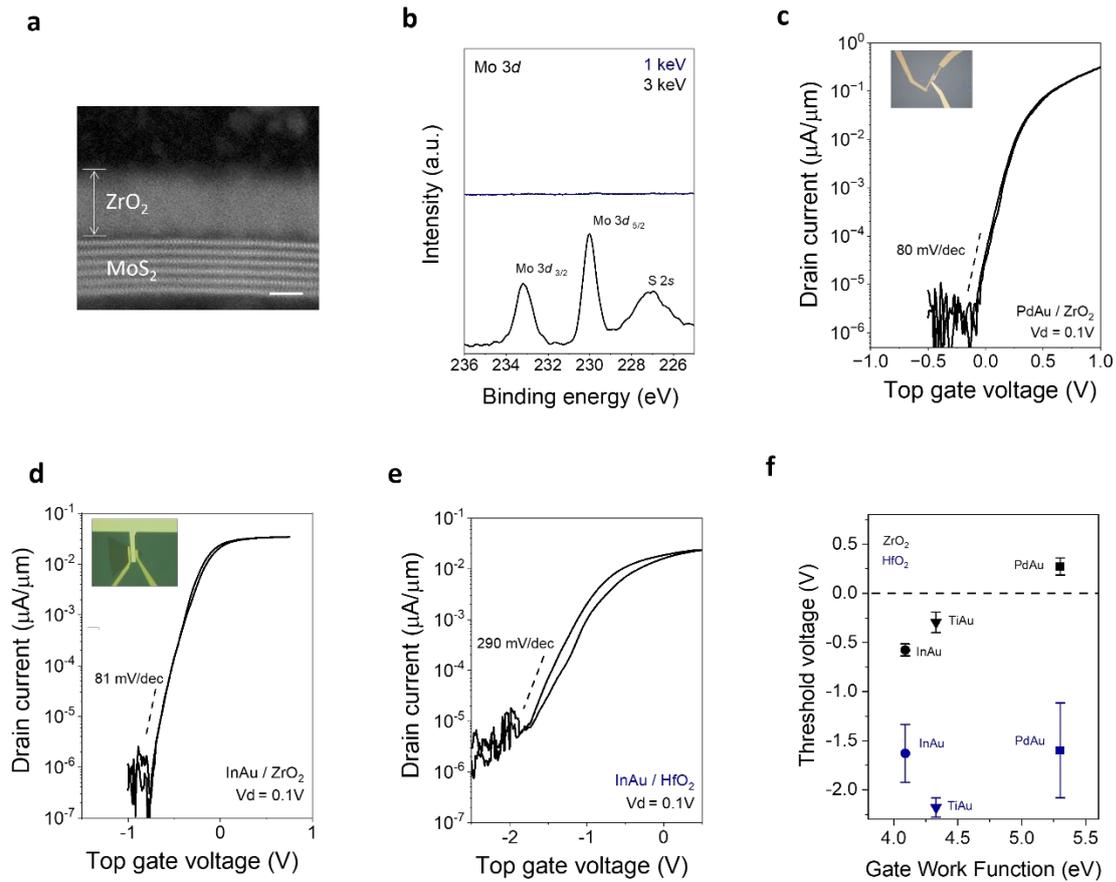

**Fig. 4. Top gate device characteristics. a,** Cross-sectional STEM of $ZrO_2$ deposited on multilayer $MoS_2$ showing defect free interface between dielectric and semiconductor. Scale bar is 2 nm. **b,** Mo 3$d$ spectra measured with 1 keV and 3keV X-rays, showing no detectable chemical interactions at the interface. 1 keV spectra in blue does not show any Mo 3$d$ signal as the X-rays only probe the top of $ZrO_2$. The 3 keV spectra in black shows Mo 3$d$ spectra that is comparable to pristine $MoS_2$ – indicating that deposition of $ZrO_2$ has no chemical or structural impact on the $MoS_2$. **c,** Typical transfer characteristics of top gate FETs based on multilayer $MoS_2$, $ZrO_2$ as the top gate dielectric, and PdAu as top gate metals. **d,** Typical transfer characteristics of FETs based on monolayer $MoS_2$, $ZrO_2$ as the top gate dielectric, and InAu as top gate metals. **e,** Typical transfer characteristics of FETs based on monolayer $MoS_2$, $HfO_2$ as the top gate dielectric, and InAu as top gate metals. **f,** $V_{th}$ of FETs based on monolayer $MoS_2$ with $ZrO_2$ and $HfO_2$ top gate dielectrics and gate metals with different work functions.

It can be seen that the $V_{th}$ can be modulated with $ZrO_2$ dielectric because of the defect free interface with $MoS_2$.

In summary, we have demonstrated that doping of $MoS_2$ is completely suppressed when $ZrO_2$ is used as the high *k* dielectric substrate. The $MoS_2/ZrO_2$ interface is ultra-clean so that high performance FETs featuring sub-1-nm EOT dielectrics, operating in enhancement mode with positive and small threshold voltage, low SS, and high ON state currents can be realized. Our results provide insight into industrially compatible dielectrics for electronics based on atomically thin semiconductors.

**Acknowledgements** M.C. and Y.W. received funding from the European Research Council (ERC) Advanced Grant under the European Union's Horizon 2020 research and innovation programme (grant agreement GA 101019828-2D- LOTTO), EPSRC (EP/ T026200/1, EP/T001038/1, EP/Z535680/1), Department of Science, Innovation and Technology and the Royal Academy of Engineering under the Chair in Emerging Technologies programme. H.Y.J. acknowledges support from the National Research Foundation of Korea (NRF) grant funded by the Korea government (MSIT) (2022R1A2C2011109). We acknowledge Diamond Light Source for time on beamline I09 under Proposal SI30105-1, SI33391-1, SI32963-1, and SI38086-1.

**Author Contributions** MC and YW supervised the project and wrote the paper. HY synthesized all the samples, measured all devices and analysed the results with YW. YL and LL assisted in device fabrication. HY, YW and DP did the synchrotron measurements with the help from TL. HG and YG did the calculation. MHK and HYJ did the STEM measurements. All authors commented on the MS.

**Methods:**

**Sample preparations:** Monolayer $MoS_2$ was obtained by CVD growth using $MoO_3$ and sulphur as precursors. 5 mg $MoO_3$ powder was evenly distributed in an alumina boat and located at the centre of a single-zone tube furnace. $SiO_2$ substrates, spin-coated with 0.5 mg/mL NaOH promoter, were placed face-down on the $MoO_3$ boat. 60 mg sulphur powder in another alumina boat was located 17 cm upper stream at the edge of the furnace. Before starting growth, the tube was purged with 460 sccm $N_2$ for 15 minutes at 150 °C. The $N_2$ flow rate was then decreased to 60 sccm. The $MoO_3$ source was heated to 720 °C and kept for 10 minutes, while the temperature of sulphur stabilized at around 230 °C. After the growth is done, the sulphur source was pulled out of the heating zone, and the tube was cooled down in 460 sccm $N_2$ gas. Polydimethylsiloxane dry transfer technique is used to transfer chemical vapor deposition grown monolayer samples to target dielectrics (see more details in supporting information). Few-layered $MoS_2$ and $WSe_2$ samples were mechanically exfoliated from the bulk crystals (purchased from 2D Semiconductors) via the blue tape method. The target substrates are the same as mentioned above for transferred CVD samples. 90 nm of thermally grown $SiO_2$, 40 nm of ALD grown $HfO_2$, and 40 nm of ALD grown $ZrO_2$ on boron degenerately doped silicon substrates were used in this study. The deposition conditions and characterisations of the dielectrics are given in supporting information.

**Soft and Hard X-ray photoelectron spectroscopy (SXPES and HAXPES) measurements:** SXPES and HAXPES measurements were conducted at Beamline I09 at the Diamond Light Source (UK) (beamtime under Proposal SI30105-1, SI33391-1, SI32963-1 and SI38086-1). The spectra were collected through a Scienta Omicron EW4000 high-energy analyzer. The beam size was around 15 μm by 35 μm. 50 eV passing energy was used for soft X-ray (1000 eV), whereas 200 eV passing energy was used for hard X-ray (3000 eV and 5900 eV). The binding energy scale was calibrated with the Au 4*f* core level of the gold electrodes on the samples as well as a gold foil on the sample holder. Radiation check was done on all samples to confirm there was no sample charging or beam damaging during measurements. This included repeating acquisition of core level spectra five times. And we proceeded to collect the actual spectra only if these radiation check spectra did not show any shift of binding energy or change of line shape. The $MoS_2$ samples used for XPS measurements were CVD-grown and transferred onto corresponding substrates, same as those utilized in our device fabrication.

**Computational details:** First-principles DFT calculations were carried out with the projector augmented wave formalism as implemented in QuantumATK package [31]. The hybrid functional of Hyed–Scuseria–Ernzerhof (HSE06) [32] was adopted with 85 Hartree cutoff energy for the plane-wave basis. Monoclinic $HfO_2$ with optimized lattice of a=5.14 Å, b=5.20 Å and c= 5.33 Å、$ZrO_2$ with optimized lattice of a=5.18 Å, b=5.25 Å and c= 5.37 Å and βcristobalite $SiO_2$ with optimized lattice of a=b=4.95 Å and c=7.31 Å are chosen to match hexagonal $MoS_2$ monolayer with optimized lattice of a=b=3.17 Å. A 3×3×1 supercell model of $MoS_2$ and 2×2×1 supercell model of $SiO_2$ (001) surface were used to build $SiO_2/MoS_2$ interface model. While for $HfO_2/MoS_2$ and $ZrO_2/MoS_2$ interface models, 3×√3×1 supercell model of $MoS_2$ was chosen to place on 2×1×1 supercell model of $HfO_2/ZrO_2$ (001) surface. A vacuum layer thickness of 20 Å was adopted in three interface models to avoid the interactions between the imaging slab. The vdW correction with Grimme's scheme was applied to describe the interface interactions [33]. The k-point samplings of 5 × 10× 1 was used for the interface structural relaxations (converged to 0.01 eV Å$^{-1}$) and 21 × 41 × 1 for the local density of states (LDOS) calculations.

**Measurements:** Electrical measurements were carried using the Keithley 4200 current voltage system. PL and Raman data was collected using a 514-nm laser excitation focused through a ×

100 objective lens. The spectra were taken at an incident laser power of 50 µW, which was sufficiently low to avoid any damage to the sample. AFM data were obtained by Dimension Icon (Bruker) with peak-force tapping (ScanAsyst) mode.

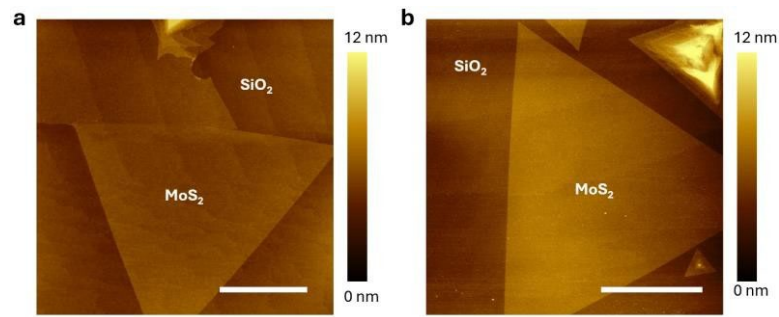

**Extended Data Fig. 1**. **Clean transfer of CVD grown MoS$_2$. a** and **b**, Transferred CVD grown MoS$_2$ flakes on SiO$_2$ using UV Ozone cleaned PDMS. Scale bar = 10 µm.

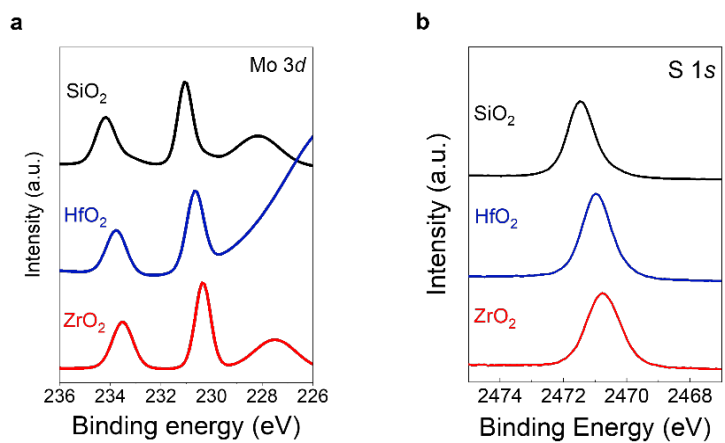

**Extended Data Fig. 2**. **Hard X-ray photoelectron spectroscopies. a**, XPS of monolayer MoS$_2$ transferred on different substrate indicates more electron doping of MoS$_2$ on SiO$_2$ compared to HfO$_2$ and ZrO$_2$ collected. Blue curve shows an obvious Hf 4$d$ peak at low binding energy due to higher probing depth. **b**, S 1$s$ spectra collected with 3keV beam energy also show higher binding energy for MoS$_2$ on SiO$_2$ compared to HfO$_2$ and ZrO$_2$.

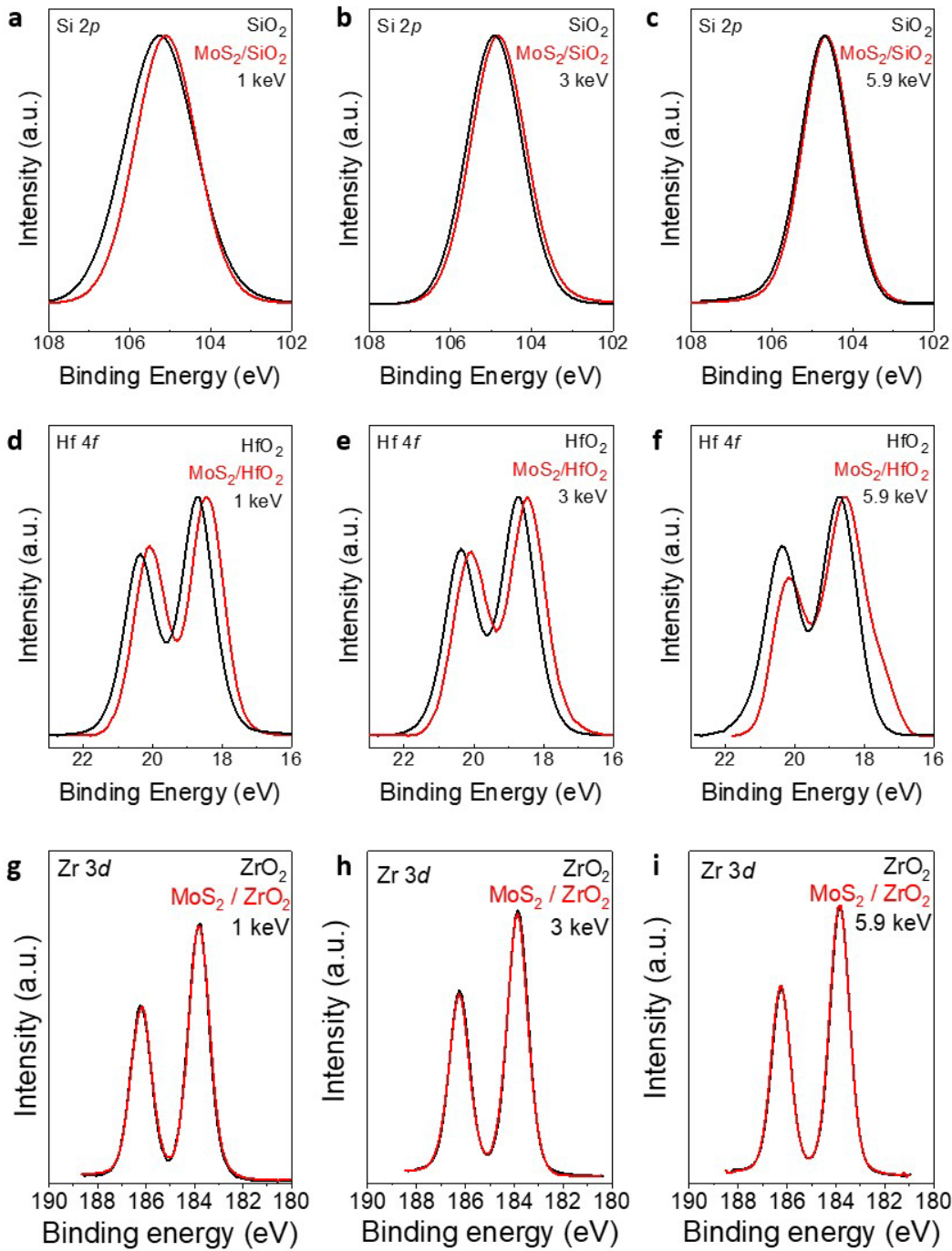

**Extended Data Fig. 3. XPS of dielectrics and MoS₂/dielectrics with different probing energy. a - c,** Si 2*p* spectra of SiO$_2$ and MoS$_2$/SiO$_2$ probed with 1 keV, 3 keV, and 5.9 keV, showing a higher binding energy in pristine SiO$_2$ compared to SiO$_2$ with MoS$_2$ on top. This shift suggests that SiO$_2$ induces n-type doping in MoS$_2$. **d - f,** Hf 4*f* of HfO$_2$ and MoS$_2$/HfO$_2$ probed with 1 keV, 3 keV, and 5.9 keV, similarly showing a higher binding energy in pristine

HfO$_2$ than in HfO$_2$ with MoS$_2$ on top, indicating n-type doping of MoS$_2$ by HfO$_2$. **g - i**, Zr 3*d* of ZrO$_2$ and MoS$_2$/ZrO$_2$ probed with 1 keV, 3 keV, and 5.9 keV showing no measurable differences across all energies, suggesting minimal electronic interaction between MoS$_2$ and ZrO$_2$.

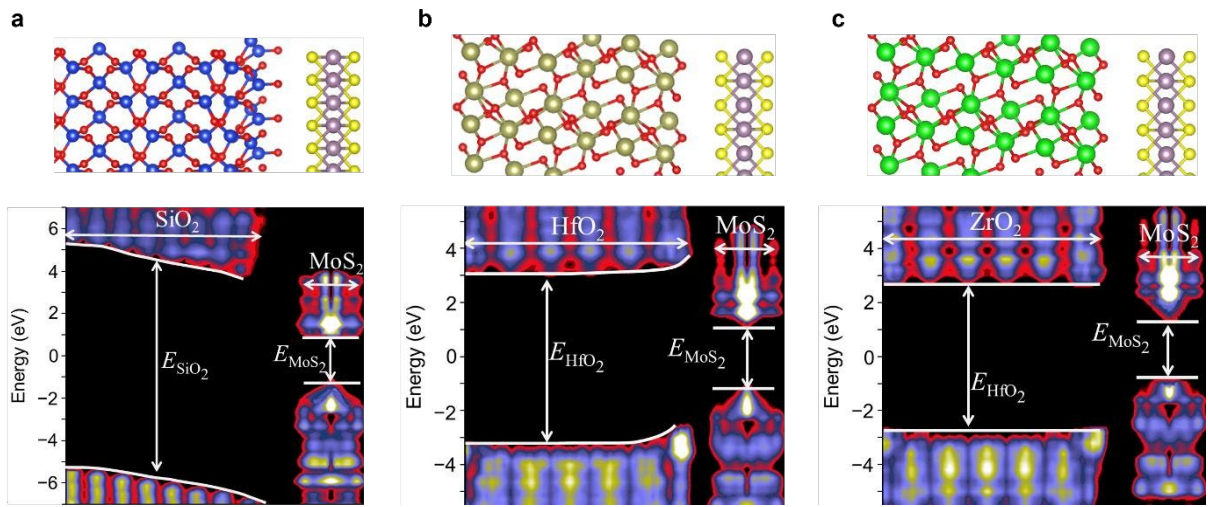

**Extended Data Fig. 4. Calculated local density of states. a,** Relaxed ball and stick model structures of SiO$_2$ (silicon = blue and oxygen = red) and MoS$_2$ (purple = Mo, yellow = sulphur) on top. LDOS for MoS$_2$/SiO$_2$ interface. E$_{SiO_2}$ is the band gap energy of SiO$_2$. The valence and conduction bands bend downward due to interface dipole formed by unpassivated oxygen atoms. **b**, Relaxed ball and stick model structures of HfO$_2$ (hafnium = gold and oxygen = red) and MoS$_2$ (purple = Mo, yellow = sulphur) on top. The HfO$_2$ surface reconstructs to form bridging oxygens. This leads to high electron density indicated by the bright white spot at valence band maximum near the surface. The states created by these bridging oxygens are at higher energy than the Hf-O bonds which causes the bands to bend upwards. **c**, Relaxed ball and stick model structures of ZrO$_2$ (zirconium = green and oxygen = red) and MoS$_2$ (purple = Mo, yellow = sulphur) on top. While bridging oxygens create high electron density at the valence maximum, their energies are comparable to the valence band energy so no band bending is observed. LDOS for MoS$_2$/ZrO$_2$ interface showing no band bending at the interface.

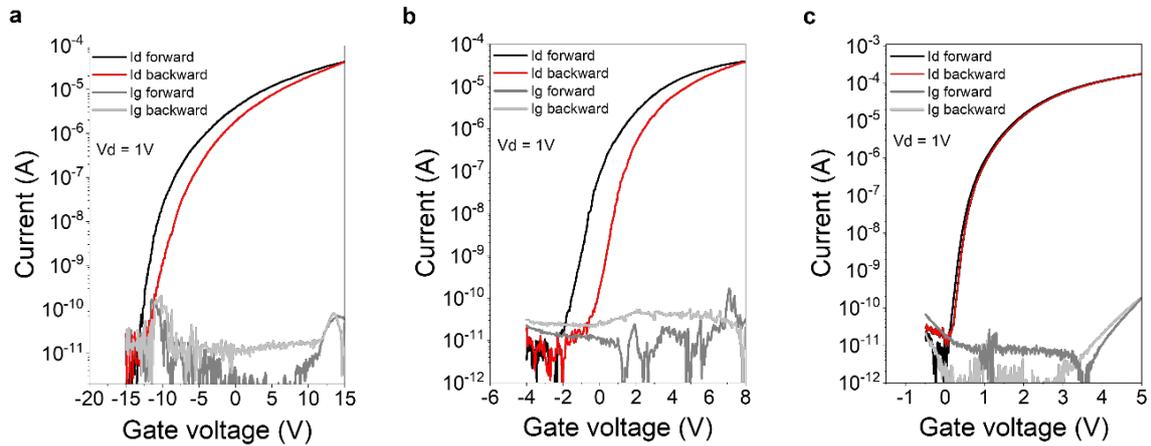

**Extended Data Fig. 5. Hysteresis in monolayer MoS$_2$ transfer curve.** Forward (black) and backward (red) scans of MoS$_2$ FETs with (**a**) SiO$_2$, (**b**) HfO$_2$, and (**c**) ZrO$_2$ back gate dielectrics. It can be seen that the hysteresis of monolayer MoS$_2$ on ZrO$_2$ is much lower compared to SiO$_2$ and HfO$_2$ due to low interface defect states.

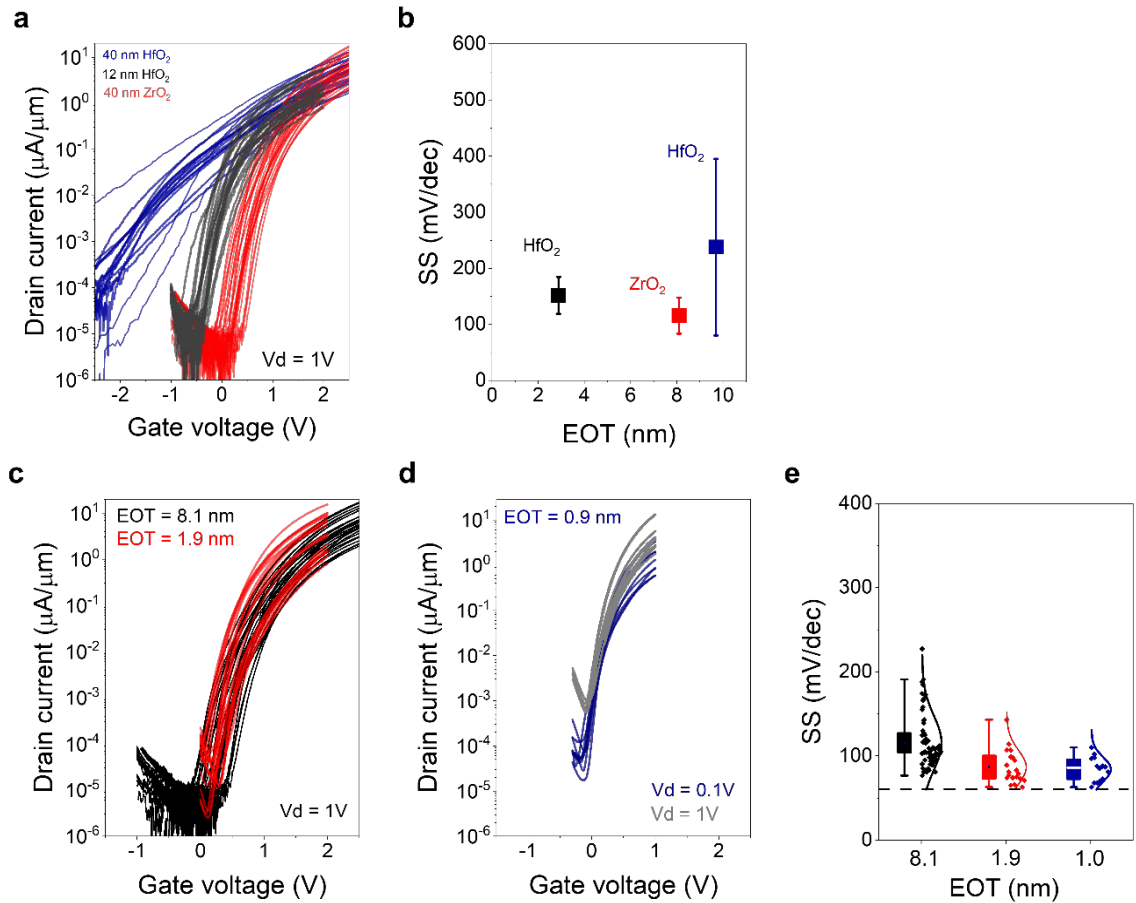

**Extended Data Fig. 6.** Thin dielectrics of $HfO_2$ and $ZrO_2$. **a**, Transfer curves of monolayer $MoS_2$ FETs on 40 nm $HfO_2$, 12 nm $HfO_2$ and 40 nm $ZrO_2$ dielectrics, respectively. $MoS_2$ FETs with $HfO_2$ dielectrics (both high and low EOTs) show negative threshold voltages. **b**, Comparison of SS values for $HfO_2$ and $ZrO_2$ dielectrics show that $MoS_2$ FETs with $ZrO_2$ dielectric exhibit low SS even with high EOT. **c**, Transfer curves of monolayer $MoS_2$ FETs with $ZrO_2$ dielectric for EOTs of 8.1 nm and 1.9 nm. **d**, Transfer curves of monolayer $MoS_2$ FETs with $ZrO_2$ dielectric for EOT of 0.9 nm at different source-drain biases. **e**, Statistical distribution of SS values for monolayer $MoS_2$ FETs with $ZrO_2$ dielectrics.

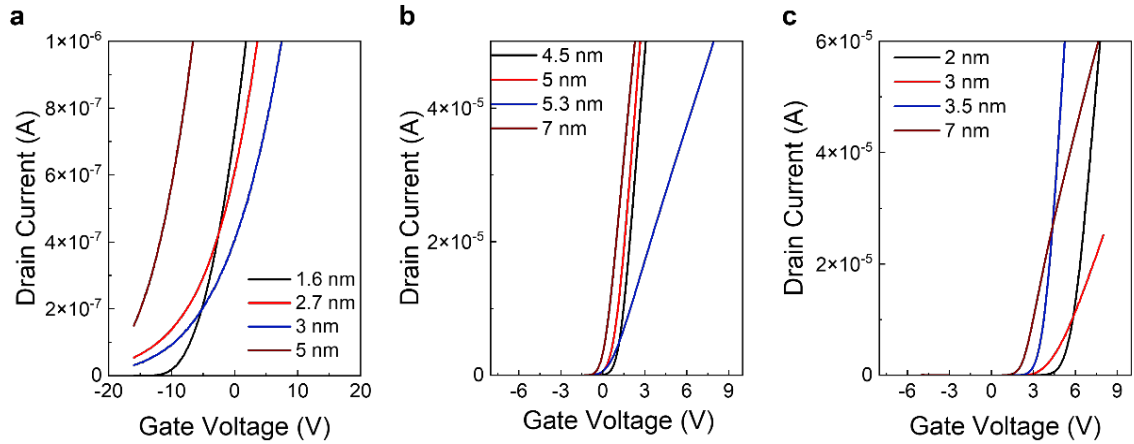

**Extended Data Fig. 7. Device characteristics of multilayer MoS$_2$ FETs on different dielectrics. a**, Transfer curve of multilayer MoS$_2$ FETs with SiO$_2$ shows negative threshold voltage. **b**, Transfer curve of multilayer MoS$_2$ FETs with HfO$_2$ shows threshold voltage close to zero. **c**, Transfer curve of multilayer MoS$_2$ FETs with ZrO$_2$ shows positive threshold voltage.

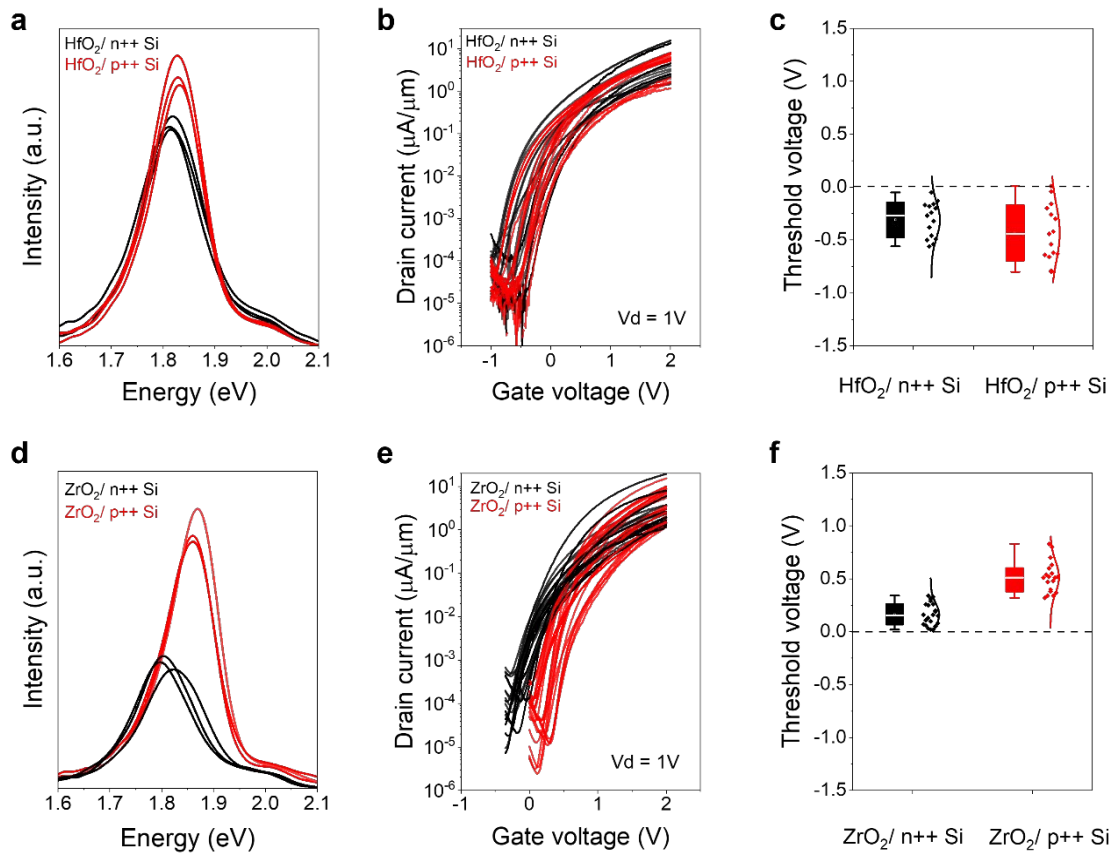

**Extended Data Fig. 8**. **Influence of back gate work function on monolayer MoS$_2$. a**, PL spectra of CVD grown monolayer MoS$_2$ transferred onto HfO$_2$ grown on n++ Si and p++ Si. **b**, Transfer characteristics of monolayer MoS$_2$ FETs with HfO$_2$ dielectrics deposited n++ Si and p++ Si, showing minimal variation despite differences in gate electrode work function. **c**, Threshold voltages extracted at 10 nA/μm for MoS$_2$ FETs with HfO$_2$ dielectrics, comparing n++ Si and p++ Si gate electrodes. **d**, PL spectra of CVD grown monolayer MoS$_2$ transferred onto ZrO$_2$ with different gate electrodes. Samples with n++ Si show trion dominant peak while p++ Si show more exciton dominant characteristics. **e**, Transfer characteristics of monolayer MoS$_2$ FETs with different gate electrodes, incorporating ZrO$_2$ dielectric layers. A lower threshold voltage is observed for metals with a lower work function. **f**, Threshold voltages extracted at 10 nA/μm for MoS$_2$ FETs with ZrO$_2$ dielectrics, comparing n++ Si and p++ Si gate 506 electrodes.

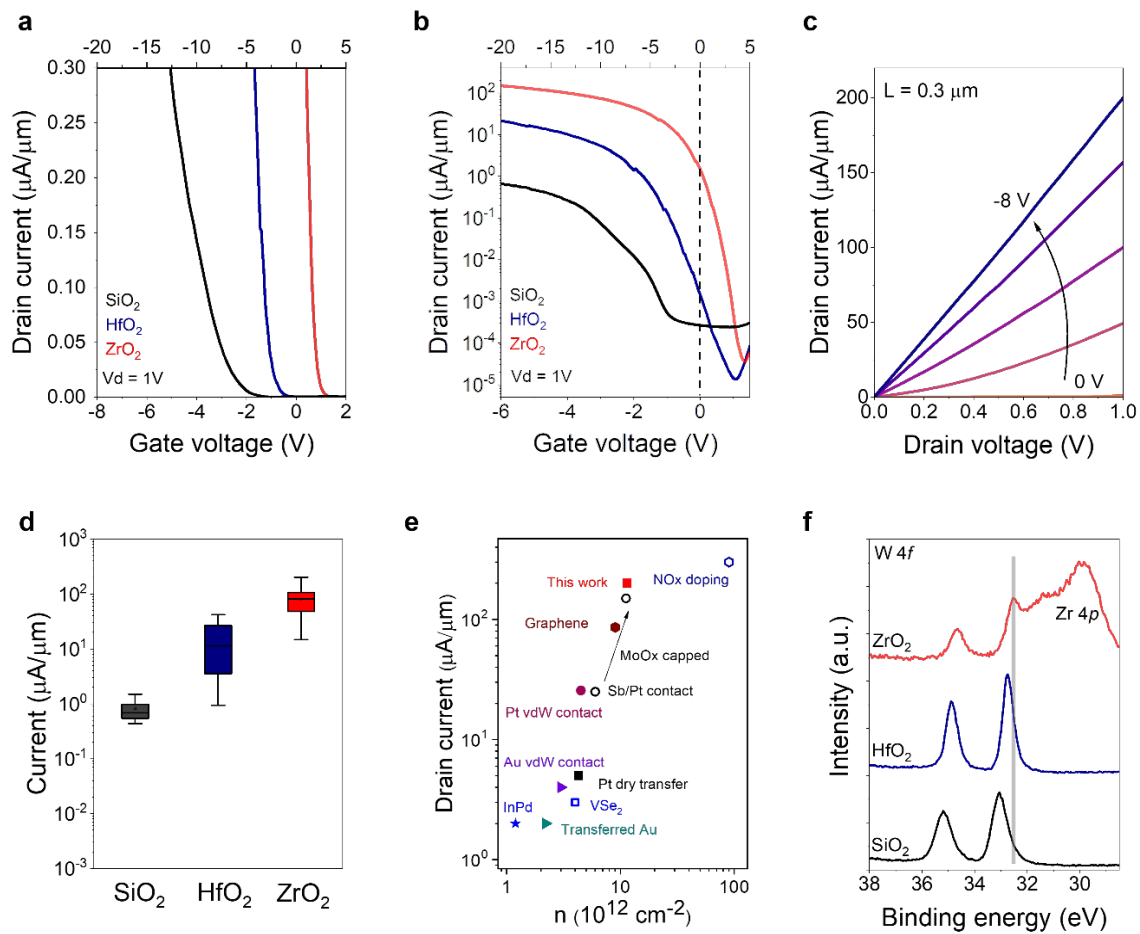

**Extended Data Fig. 9. WSe$_2$ P-type FETs with SiO$_2$, HfO$_2$ and ZrO$_2$ dielectrics. a,** Transfer curves of WSe$_2$ FETs plotted linearly to show the negative threshold voltages for WSe$_2$ FETs with SiO$_2$ and HfO$_2$ and positive threshold voltage for WSe$_2$ FETs with ZrO$_2$. The gate voltage scan range for SiO$_2$ is shown on the top axis. **b,** Logarithmic transfer characteristics of WSe$_2$ FETs with SiO$_2$, HfO$_2$ and ZrO$_2$, highest hole current of 201 µA/µm for WSe$_2$ on ZrO$_2$. The gate voltage scan range for SiO$_2$ is shown on the top axis. **c,** Output curves of WSe$_2$ FETs with ZrO$_2$ showing linear current-voltage characteristics. **d,** Hole current distribution of >20 WSe$_2$ FETs on different dielectrics. ZrO$_2$ dielectric shows the highest hole current compared to HfO$_2$ and SiO$_2$. **e,** Drain current of P-type WSe$_2$ FETs with values reported in the literature [8,34–42]. Solid symbols are for multilayer WSe$_2$ FETs and empty symbols are for monolayer WSe$_2$ FETs.

**f,** W 4*f* core level spectra of WSe$_2$ on SiO$_2$, HfO$_2$, and ZrO$_2$. The higher binding energy of W 519 4*f* from WSe$_2$ on SiO$_2$ indicates higher electron doping of WSe$_2$ from the dielectric.

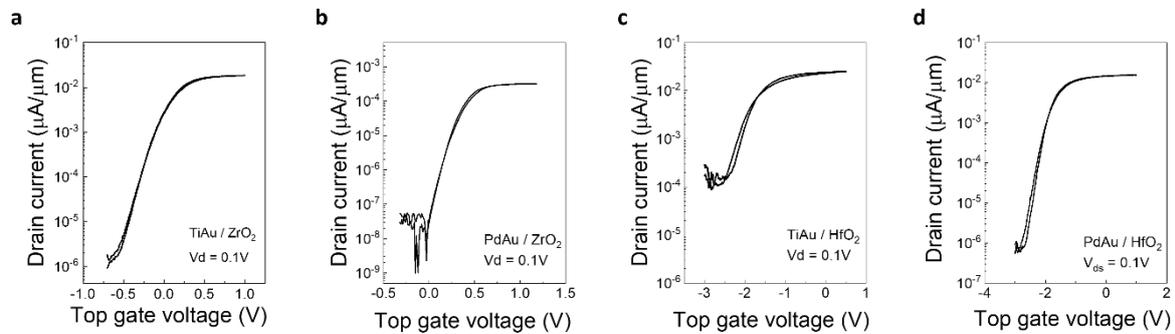

**Extended Data Fig. 10**. **a,** Typical transfer characteristics of FETs based on monolayer MoS$_2$, ZrO$_2$ as the top gate dielectric and TiAu as top gate metals. **b,** Typical transfer characteristics of FETs based on monolayer MoS$_2$, ZrO$_2$ as the top gate dielectric and PdAu as top gate metals. **c,** Typical transfer characteristics FETs based on monolayer MoS$_2$, HfO$_2$ as the top gate dielectric and TiAu as top gate metals. **d,** Typical transfer characteristics of FETs based on monolayer MoS$_2$, HfO$_2$ as the top gate dielectric and PdAu as top gate metals.